\documentclass[12pt,preprint]{aastex}

\shorttitle{O~{\sc ii} ground configuration energy levels}
\shortauthors{Blagrave and Martin}

\begin{document}

\title{On the O~{\sc ii} ground configuration energy levels}

\author{K. P. M. Blagrave and  P. G. Martin}
\affil{Department of 
Astronomy and Astrophysics and Canadian 
Institute for Theoretical Astrophysics, University of Toronto,
60 St. George Street, Toronto, ON M5S 3H8, Canada}
\email{blagrave@cita.utoronto.ca, pgmartin@cita.utoronto.ca}

\begin{abstract}

The most accurate way to measure the energy levels for the O~{\sc ii} $2p^3$
ground configuration has been from the
forbidden lines in planetary nebulae.  We present an analysis of
modern planetary nebula data that nicely constrain the splitting
within the $^2$D term and the separation of this term from the ground
$^4$S$_{3/2}$ level.
We extend this method to H~{\sc ii} regions using high-resolution
spectroscopy of the Orion nebula, covering all six visible transitions
within the ground configuration.  These data confirm the splitting of
the $^2$D term while additionally constraining the splitting of the
$^2$P term.  The energies of the $^2$P and $^2$D terms relative to the
ground ($^4$S) term are constrained by requiring that all six lines
give the same radial velocity, consistent with independent limits
placed on the motion of the O$^+$ gas and the planetary nebula data.
\end{abstract}


\section{Introduction \label{int}}
\begin{figure}
\epsscale{0.9}
\plotone{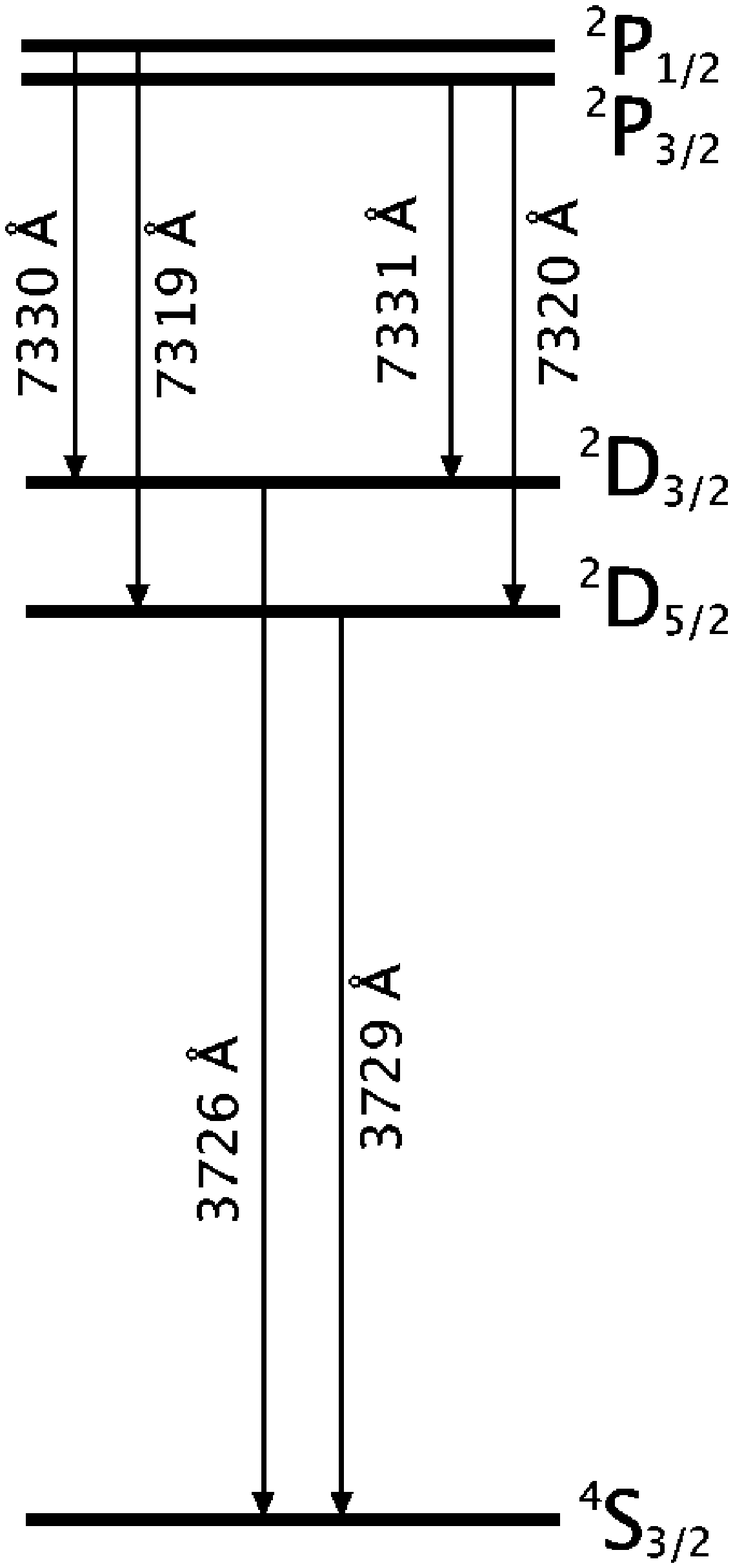}
\caption{Grotrian diagram of O~{\sc ii} ground configuration energy level
transitions (not to scale).  Wavelengths are approximate air values. 
\label{figene}} \end{figure}

In this paper we determine the four energies which describe the
separation of the five ground configuration energy levels of O~{\sc
ii} shown in Figure~\ref{figene}, through analysis of observations of
the visible forbidden transitions.  Being forbidden transitions, these
are weak and not accessible in the laboratory.  The ground
configuration levels can be determined from laboratory data on
permitted ultraviolet transitions from higher states, but in practice
the uncertainties in the derived energy levels are larger than those
obtained from the nebular visible transitions.  Thus the published
compilations of data on these energy levels (Eriksson 1988; Martin, Kaufman
\& Musgrove 1993)
rely heavily on the pioneering work of \citet{bow60} and de Robertis,
Osterbrock, \& McKee (1985)
on planetary nebulae (see Table~\ref{tbl1}).  Our analysis of new data
improves upon these energies and provides a discussion of the
uncertainties.

\citet{bow60} observed the two blue [O~{\sc ii}] lines in seven bright
planetary nebulae (a total of sixteen photographic plates) and the
four red [O~{\sc ii}] lines (blended into two lines) in one nebula
(three separate plates).  The O$^+$ gas of the planetary nebulae was
assumed to have an average bulk velocity identical to the measured
velocity of the H$^+$ gas (the H$^+$ energy levels being well known).
The observed [O~{\sc ii}] wavelengths adjusted to this velocity give
absolute wavenumbers for the energy level differences for the
transitions.  These early data resolved the $^2$D term into its two
$J$ components, but were not able to determine the splitting of the
$^2$P term.
The latter splitting was determined in \citet{der85} using
high-resolution digital spectra of the four red lines in one bright
planetary nebula (NGC 7027).  As can be appreciated from
Figure~\ref{figene}, the red transitions ending on a common lower level
constrain the splitting of the $^2$P term while the transitions from a
common upper level constrain the splitting of the $^2$D term,
independently of the blue data.  \cite{der85} found a value for the
latter splitting somewhat different than that found by Bowen.
Since absolute velocities or wavenumbers were not measured,
\cite{der85} calculated the $^2$D--$^2$P separation by assuming that
the $\lambda$7319/$\lambda$7320 line blend measured by \citet{bow60}
was actually a measure of the stronger $\lambda$7320 line, and that
\citet{bow60}'s roughly symmetric $\lambda$7330/$\lambda$7331 blend
was a measure of this line pair's average.

We show that H~{\sc ii} regions can also be used to determine the
energy levels of O~{\sc ii}.  With high-resolution echelle
spectroscopy of the Orion nebula (resolving all red and blue lines),
we are able to measure not only the splitting of both the $^2$P and
$^2$D terms but also the separation between these two split terms.
The observations are described in \S~\ref{obs} and the full analysis,
including uncertainties, in \S~\ref{con}.  But first we update Bowen's
work by analysis of more recent digital data on 23 planetary nebulae
(\S~\ref{bow}).  This provides important insight into the method of
analysis of the full Orion data and also provides improved values of the two
energies constrained by the blue lines \`a la Bowen.

We conclude by comparing our results with recent work by \citet{sha04}.

\section{Bowen revisited \label{bow}}

Using a coud\'{e} spectrograph on the Hale telescope, \citet{bow55}
observed the blue spectra of seven planetary nebulae photographically.
Since the introduction of the Hamilton Echelle Spectrograph in 1987
\citep{vog87}, high-resolution spectra of (at least) 24 planetary
nebulae spectra have been published, the relevant data for our
purposes being the line lists with accurately tabulated wavelengths.
(One nebula, NGC 6818, had to be discarded because of a grossly
discrepant wavelength -- we suspect a typographical error.)  These new
spectra resolve the blue lines but not the red pairs of lines.

Following Bowen we therefore have two lines to constrain two energies.
But there is an important requirement: we need to know the velocity of
the line-producing gas and any uncertainty ($\delta v_{\mathrm{O^+}}$)
will have predictable consequences. For example, if one uses the
3729~\AA\ line to find the energy difference for this transition,
$E_{^4\mathrm{S}_{3/2}-^2\mathrm{D}_{5/2}}$ (note the convention of an
``upwards'' transition), because of this degeneracy the Doppler effect
gives an {\it uncertainty}
\begin{equation}
\label{dop}
\delta E_{^4\mathrm{S}_{3/2}-^2\mathrm{D}_{5/2}} = 0.0895 \delta v_{\mathrm{O^+}},
\end{equation}
where throughout this paper energies are in units of cm$^{-1}$ and 
velocities in km s$^{-1}$. Looking ahead to the red data, 
\begin{equation}
\label{dop2}
\delta E_{^2\mathrm{D}_{3/2}-^2\mathrm{P}_{3/2}} = 0.0455 \delta v_{\mathrm{O^+}} 
= 0.50865 \delta E_{^4\mathrm{S}_{3/2}-^2\mathrm{D}_{5/2}},
\end{equation}
but note that the smaller splittings
$E_{^2\mathrm{D}_{5/2}-^2\mathrm{D}_{3/2}}$ and
$E_{^2\mathrm{P}_{3/2}-^2\mathrm{P}_{1/2}}$ are much less susceptible
to any uncertainty in the velocity.

In Bowen's work, permitted lines of H$^+$ and He$^+$ gas set the
rest-frame velocity and similarly here all wavelengths for all 23
planetary nebulae are first put into such a reference frame with H$^+$
at rest.  As tabulated by the authors, the planetary nebulae
wavelengths have in fact already been corrected for the
previously-known systemic velocity of the whole nebula, but
nevertheless we find from the data that the H$^+$ is apparently not
quite at rest.  This is either because the systemic velocity used was
only approximate, or because for the slit positions observed the mean
velocity of the H$^+$ gas is not identical to that averaged over the
nebula, which would not be surprising given incomplete coverage of an
expanding nebula.  To define the H$^+$ frame for each nebula, we used
the eight or nine unblended H~{\sc i} lines (H16 to H$\delta$) near
the blue [O {\sc ii}] lines, these all being contained in the same
echelle spectrum.  The corresponding blue wavelengths are given in
Table~\ref{tablepn}.

The measured velocity of O$^+$ is not necessarily the same as that of 
H$^+$ because of the ionization structure and expansion that exist in  
the nebula and the fact that the slit does not usually cover the 
entire nebula.  Bowen's 
approach was to average results over 
several nebulae, since on average the two velocities should be equal.
Given data on 23 nebulae, we
can improve upon this iteratively as explained below.

Our analysis is basically to develop a parameterized model, and then
optimize the parameters by non-linear least squares to match predicted
wavelengths, in air, with the wavelengths tabulated.  The parameters
of the model are the energies and any velocity offsets ($v_{\mathrm{O}^+}$)
deemed necessary.  (Toward this end, deviations of observed wavelengths
from the model predictions are expressed in terms of velocity.)  The
energies can be taken as the successive energy differences, four
independent values in the full model, or simply the energies of the
four upper levels.  Even with the first of these two options there is
covariance in
the resulting solution, since four of the six energy transitions
(corresponding to
four of the six available wavelengths) couple the independent successive energy
differences.  The ``model uncertainties'' from the goodness of fit to 
the model are the 68.3\% confidence
intervals for one-dimensional marginal distributions for each of the
parameters.

Let us return to the blue lines, for which we have 46 measured
wavelengths for 23 planetary nebulae.  As energy parameters, we used
$E_{^4\mathrm{S}_{3/2}-^2\mathrm{D}_{5/2}}$ and
$E_{^2\mathrm{D}_{5/2}-^2\mathrm{D}_{3/2}}$.  In the initial model we
went to the extreme of introducing 23 velocity offsets.  This
precludes determining $E_{^4\mathrm{S}_{3/2}-^2\mathrm{D}_{5/2}}$ and
we find 
$E_{^2\mathrm{D}_{5/2}-^2\mathrm{D}_{3/2}} = 19.79 \pm
0.05$.  For each nebula the two velocity
residuals (from the two wavelength residuals) are of equal magnitude
(denoted $\sigma_{int}$) with opposite signs, indicating that relative
to the model the two lines are too close together or too widely
separated.  Overall the rms velocity residual was 0.96~km~s$^{-1}$.
Those nebulae with considerably larger rms values can be judged
to have data of lower quality; i.e., even with the luxury of the
maximal number of parameters, the data are still not going to be well matched by
the model.  Two nebulae with residuals greater than 2.4~km~s$^{-1}$
have been assigned weight 0.25 (in the calculation of $\chi^2$) while
another four with residuals greater than 1.2~km~s$^{-1}$ have been
assigned weight 0.5.  No bias is introduced in subsequent calculations
of the splitting $E_{^2\mathrm{D}_{5/2}-^2\mathrm{D}_{3/2}}$ since
equal numbers of ``too close'' and ``too separated'' cases are involved;
now $E_{^2\mathrm{D}_{5/2}-^2\mathrm{D}_{3/2}} = 19.79 \pm
0.04$.  The weighted rms residual is 0.71~km~s$^{-1}$ and no
subsequent model, with fewer parameters, can improve upon this.

The next model goes to the other extreme, fitting only the two
energy differences, finding 
$E_{^4\mathrm{S}_{3/2}-^2\mathrm{D}_{5/2}} = 26810.68\pm0.13$ and
$E_{^2\mathrm{D}_{5/2}-^2\mathrm{D}_{3/2}} = 19.79\pm0.18$.
The latter energy difference is still close to that in the initial model, but is
determined with less confidence because the two-parameter model fits
the data less well.  The rms velocity residual is 6.7~km~s$^{-1}$,
much larger than suggested by our assessment of the data quality, and
so clearly indicating a less than optimal model.

Thus our goal was to improve the model iteratively by adding a minimal
number of parameters $v_{\mathrm{O}^+}$, for a subset of the nebulae, 
expecting a significant reduction in $\chi^2$ per degree of freedom.  We
identified seven nebulae for which the residuals exceeded
6.7~km~s$^{-1}$ and also $3\sigma_{int} < 6.7$~km~s$^{-1}$ and for
these included a velocity offset parameter.  This produced a markedly improved
model with an rms residual of 1.9~km~s$^{-1}$ and
$E_{^4\mathrm{S}_{3/2}-^2\mathrm{D}_{5/2}} =
26810.77\pm0.04$ and
$E_{^2\mathrm{D}_{5/2}-^2\mathrm{D}_{3/2}} = 19.79\pm0.06$.
Repeating this process with the new rms identifies eight more nebulae
which would benefit from velocity offsets, for a total of 15 of the 23
nebulae.  This iteration reduces the rms velocity residual to
0.82~km~s$^{-1}$; this is now comparable to the above estimate of the
quality of the data, indicating that adding further parameters would
not be justified.
$E_{^4\mathrm{S}_{3/2}-^2\mathrm{D}_{5/2}} =
26810.77\pm0.03$ and
$E_{^2\mathrm{D}_{5/2}-^2\mathrm{D}_{3/2}} = 19.79\pm0.03$ --
the only difference with respect to the previous iteration being a
lowering of the error (confidence interval).
It is important to acknowledge that there is a 
systematic error in $E_{^4\mathrm{S}_{3/2}-^2\mathrm{D}_{5/2}}$ of order
0.03 cm$^{-1}$ because the introduction of parameters $v_{\mathrm{O}^+}$, 
while hopefully unbiased, is still subjective.
Such systematic errors are recorded separately in 
Table~\ref{tbl1} to distinguish them from confidence intervals.

The energies determined are close to those given by Bowen,
but significantly different than those derived by
\cite{der85} -- refer to Table~\ref{tbl1}.

\section{Observations of the Orion Nebula}\label{obs}

High-resolution echelle spectra were obtained over the course of two
nights in 1997 (5100-7485 \AA) and 1998 (3510-5940 \AA)
with the CTIO 4-m Blanco telescope (refer to
\citet{bal00} for details).  Three different lines of sight were
observed, referred to as 1SW, x2 \citep{bal96,rub97} and 37W
\citep{bal91}.  To extract the information on individual lines, in
particular the wavelength, the data were modeled with a Gaussian
(three parameters: central wavelength, FWHM, area) and a linear
baseline (two further parameters).  The rest-frame velocity of the
H$^+$ gas along each line of sight was determined from the six
strongest unblended H~{\sc i} Balmer lines.

\begin{figure}
\epsscale{1.0}
\centerline{\rotatebox{90}{\plottwo{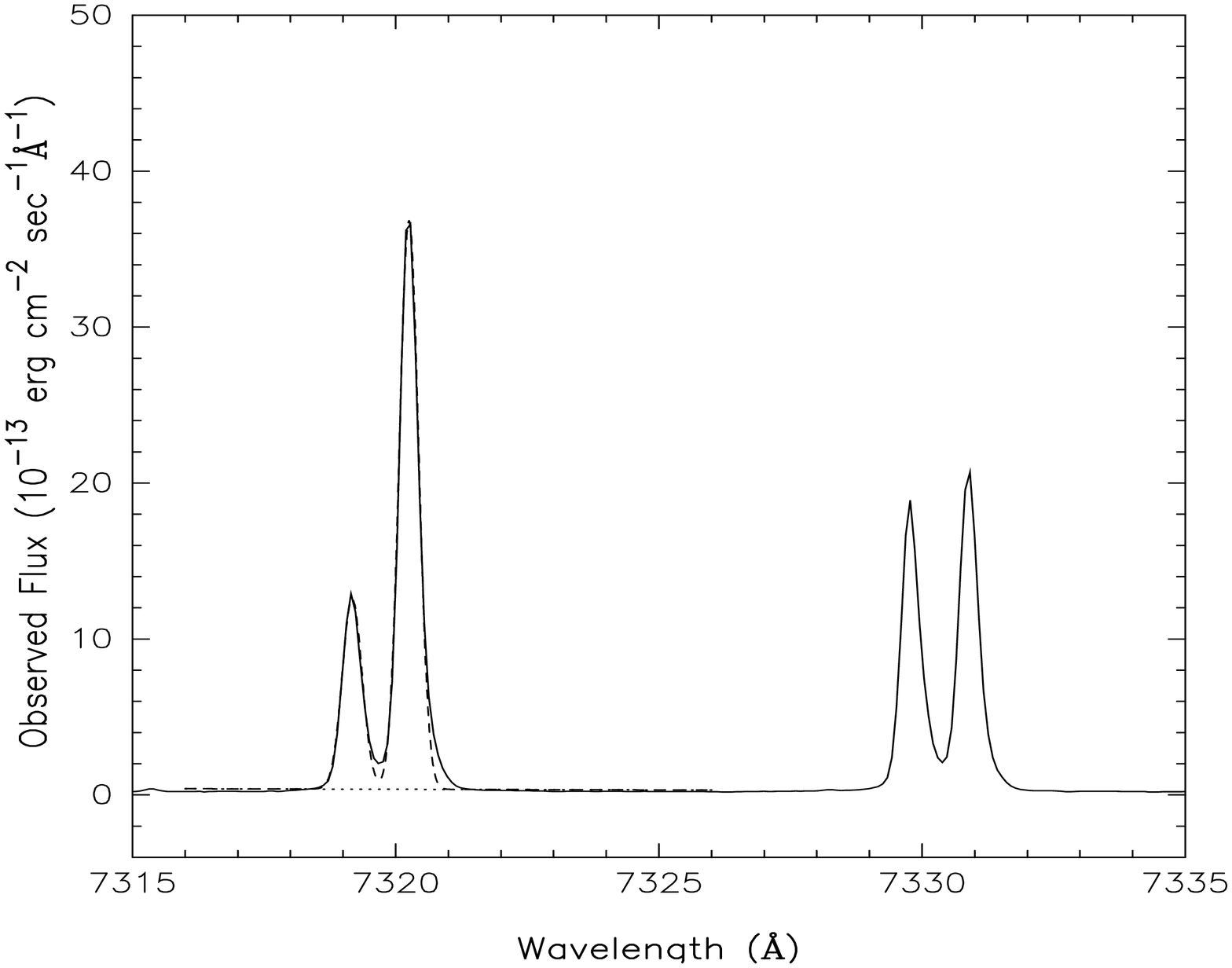}{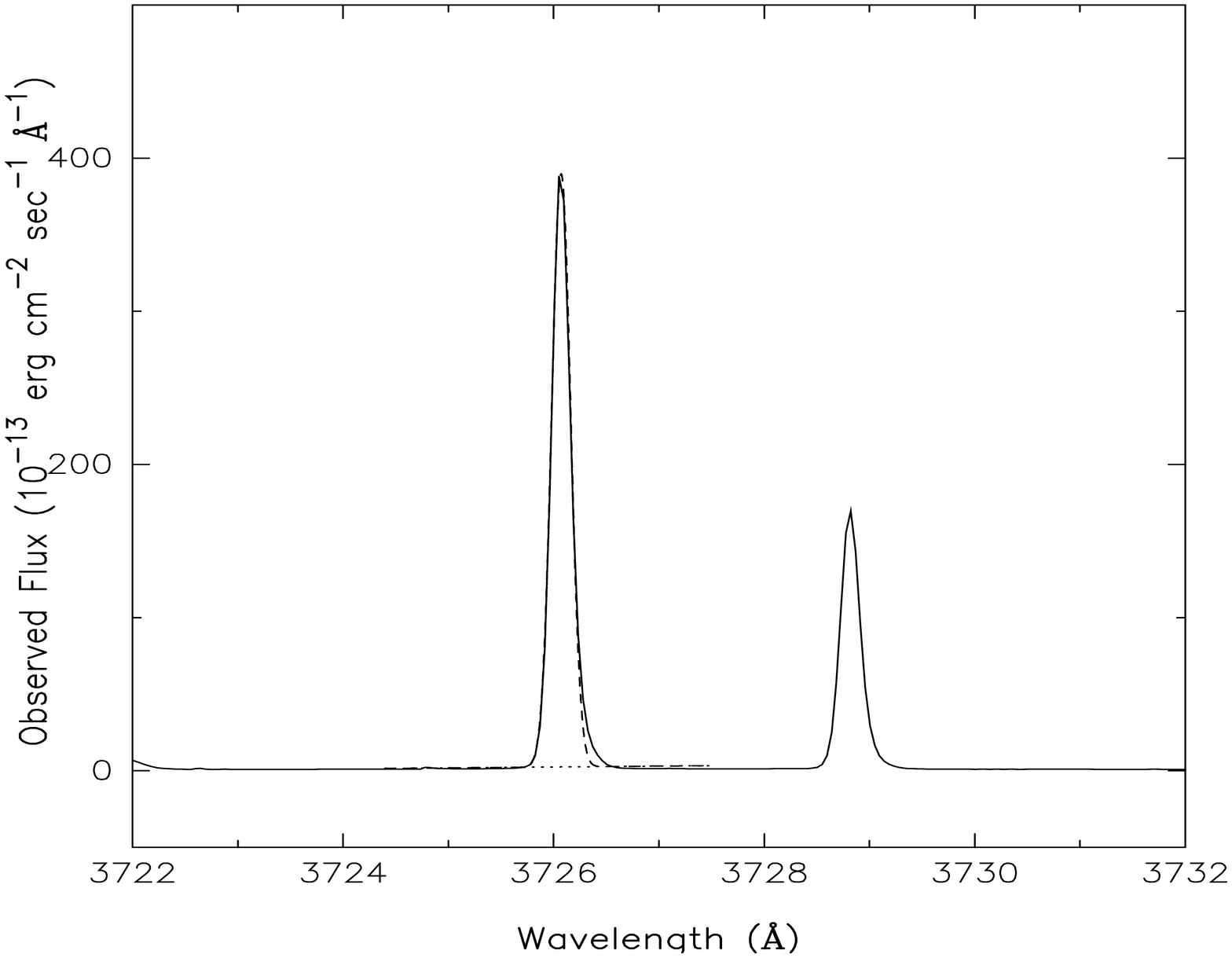}}}
\caption{Gaussian profile fit of all six visible O~{\sc ii} lines (1SW).  The
dark solid line represents the observed line profile and the lighter dashed
line represents the Gaussian fit.  The wavelength has been adjusted to the
rest-frame
velocity of the H$^+$ gas.  Refer to Table \ref{tbl2} for reddening-corrected
fluxes and the results of the line-fitting. \label{figpro}}
\end{figure}

All six [O~{\sc ii}] forbidden lines are seen with good signal to
noise in these spectra.  Data for 1SW are shown in Figure~\ref{figpro}.
The pairs of [O~{\sc ii}] red lines were slightly blended and so were
analysed using a double Gaussian fit. The results of our line fitting
are summarized in Table~\ref{tbl2}.  All line profiles are similar as
seen in the matching FWHM and directly from the spectra in
Figure~\ref{figpro}.

For the common upper level line pairs $\lambda$7320/$\lambda$7331 and
$\lambda$7319/$\lambda$7330 the line strength ratios can be predicted
directly from the transition probabilities \citep{zei87, wie96},
offering an independent check of one aspect of the fits.  Results are
presented in Table~\ref{tblrto}.  There is reasonable agreement between 
the theory and the observations.
The sole anomaly, seen in the $\lambda$7319/$\lambda$7330 ratio in the
x2 line-of-sight, arises because of a velocity-shifted component from
a photoionized Herbig-Haro shock 
\citep{bla03} that has 2-4$\%$ of the
nebular [O~{\sc ii}] flux.  The nebular $\lambda$7330 and
$\lambda$7319 lines are contaminated by the velocity-shifted
components of $\lambda$7331 and $\lambda$7320, respectively.  A higher
relative contamination from the stronger of these two lines,
$\lambda$7320, results in a higher $\lambda$7319/$\lambda$7330 ratio.

All four [O~{\sc ii}] red lines are found in the same echelle order of
a single exposure, unlike in the data of \citet{der85} where the
$\lambda$7319/$\lambda$7320 and $\lambda$7330/$\lambda$7331 line pairs
were obtained in two separate spectra.  Measuring the $^2$P energy
splitting $E_{^2\mathrm{P}_{3/2}-^2\mathrm{P}_{1/2}}$ depends on the
wavelength difference {\it within} each of the line pairs and so the
results in \citet{der85} should be accurate.  On the other hand, the
splitting of the $^2$D term
$E_{^2\mathrm{D}_{5/2}-^2\mathrm{D}_{3/2}}$ depends on the wavelength
difference {\it between} the pairs and so our single spectrum
containing both line pairs, with only a single wavelength calibration
in the same echelle order, should yield a more accurate result.

All lines produced by the same ionized species should have the same
velocity.  However, using the current published energy levels and
derived rest wavelengths in air \citep{eri87, mar93} together with our
observed wavelengths, we obtain O$^+$ velocities that are grossly
inconsistent, well beyond the uncertainties propagated from the
measurement errors of the observed wavelengths (see
Table~\ref{tblvel}).  In particular, there is no explanation why lines
in the same wavelength region and originating from a common upper
level ($^2$P$_{1/2}$ or $^2$P$_{3/2}$) should yield significantly
different velocities, as is observed to be the case in columns 7 and 8
of Table~\ref{tblvel}; at the very least, there is a problem with the
splitting $E_{^2\mathrm{D}_{5/2}-^2\mathrm{D}_{3/2}}$.  The
consistency of the data for the three lines of sights, and the lack of
agreement of the velocities from all of the lines, points to
inaccurate [O~{\sc ii}] rest wavelengths arising from poorly
determined ground configuration energy levels, including the
separations $E_{^4\mathrm{S}_{3/2}-^2\mathrm{D}_{5/2}}$ and
$E_{^2\mathrm{D}_{3/2}-^2\mathrm{P}_{3/2}}$.  This was the original
motivation for this paper.

\section{Constraining the Energy Levels with the Orion Nebula Data}\label{con}

With a set of six accurate wavelengths, for each of three
lines-of-sight (1SW, x2, and 37W), it is possible to obtain the
energies of the four excited levels in the ground configuration of
O~{\sc ii} (Fig.~\ref{figene}); the problem is over-constrained.
However, as encountered in the analysis of planetary nebula spectra in
\S~\ref{bow}, there is the possibility of an unknown velocity offset
of the O$^+$ gas for each position observed.  But even with an extra
velocity offset parameter $v_{\mathrm{O}^+}$ for each position, the 
problem is 
still over-constrained (even for a single position) and thus amenable to
modeling and least-squares optimization.

As with the planetary nebulae, there is a Doppler-related degeneracy
to be resolved as well, through independent constraints on $v_{\mathrm{O}^+}$.
We show how
this is possible in Orion, it not being sufficient to assume that on
average, over many positions, the velocities of the O$^+$ gas and
H$^+$ gas are identical (in which case $\left<v_{\mathrm{O}^+}\right> = 0$).

\begin{figure}
\epsscale{.72}
\rotatebox{270}{\plotone{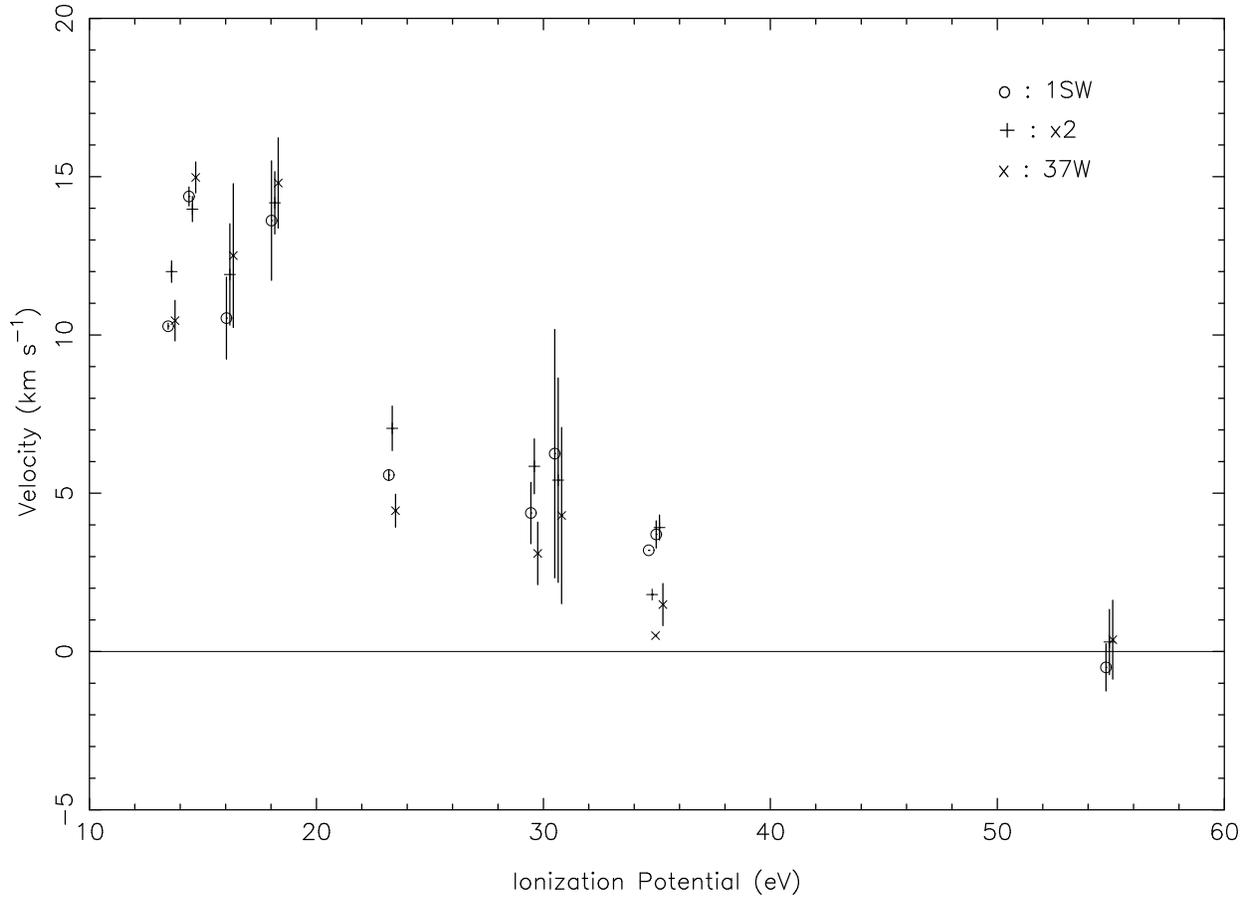}}
\caption{Gas velocity as a function of the emitting species' ionization 
potential.  The three lines-of-sight have been offset in the x-direction to
properly show the standard deviations about the mean velocities
for each line-of-sight.
 The ionization potentials of O$^0$ and O$^{++}$ are 13.6181 eV and
54.934 eV, respectively. 
O$^+$ ionization potential is 35 eV.\label{figion}}
\end{figure}

The Orion nebula can be represented by a blister model where gas is
accelerating toward the ionizing star and the observer away from the
background molecular cloud \citep{bal74}.  Because of ionization
stratification in this accelerating flow, the velocity is correlated
with the ionization potential (I.P.): gas that is more highly ionized is
more blue shifted (has more negative velocity).  This is observed;
see, e.g., \citet{bal00}.  Figure~\ref{figion}, plotting velocity
against the emitting species' I.P., summarizes this
effect for our observations of many lines of many different ions for
the three lines of sight.  Models of the nebula \citep{bal00} show
that the O$^+$ zone is relatively narrow (see their Figure 4) and so
there should be a well defined velocity.  From the trend seen in
Figure~\ref{figion}, the expected value for $v_{\mathrm{O}^+}$, at I.P. 35~eV 
for O$^+$, lies between 0 and 5 km~s$^{-1}$, set in part by the
[S~{\sc iii}] (I.P. = 34.79 eV)
velocities at similar ionization potential. Plotting the results
in columns 7 and 8 of Table~\ref{tblvel} in Figure~\ref{figion} would
clearly reveal their discrepancy, again pointing to a problem with the
energy levels. The velocities from our new O$^+$ model clearly satisfy the 
general constraint set by the surrounding lines.

As in the analysis of the planetary nebula data, our examination of the 
Orion nebula data is rooted in a model of
the energy levels, used to predict the air wavelengths.  In this case
we set $E_{^4\mathrm{S}_{3/2}-^2\mathrm{D}_{5/2}} =
26810.77$, the result already found in \S~\ref{bow}.
The remaining analysis is then a non-linear least squares fit
(unweighted), using six parameters in the model: the other three
energy differences, and the three velocities $v_{\mathrm{O}^+}$.
This model produces rms energy-difference residuals of only 0.4~km~s$^{-1}$.
Line by line and position by position, this model
gives the velocities listed in column 9 of Table~\ref{tblvel}; these 
are clearly now quite consistent with one another.

This choice of $E_{^4\mathrm{S}_{3/2}-^2\mathrm{D}_{5/2}}$ produces 
$v_{\mathrm{O}^+}$ equal to
$4.0\pm0.3$~km~s$^{-1}$, $4.1\pm0.3$~km~s$^{-1}$, and
$1.4\pm0.3$~km~s$^{-1}$ for 1SW, x2, and 37W, respectively (because
of changes in geometry from one line-of-sight to the next,
these velocities need not be identical).  It
turns out that these fall in line with the trend in Figure~\ref{figion} and so
no optimization was carried out on this energy difference.  

Recall however, that if a slight change were made in this energy,
$v_{\mathrm{O}^+}$ would respond according to 
equation~\ref{dop}.  Thus the model uncertainty $\delta
E_{^4\mathrm{S}_{3/2}-^2\mathrm{D}_{5/2}}$ from the planetary nebulae 
analysis, 
$\pm 0.03$~cm$^{-1}$, results in
$\delta v_{\mathrm{O}^+} = 0.34$~km~s$^{-1}$.  This is of the same order 
as the
model fitting uncertainties for the Orion data.
In principle, one might start by determining $v_{\mathrm{O}^+}$ and its 
uncertainty from interpolation in Figure~\ref{figion} and work backwards to 
$E_{^4\mathrm{S}_{3/2}-^2\mathrm{D}_{5/2}}$ and its uncertainty.
Given the consistency, we adopt the tighter constraint found independently 
in \S~\ref{bow}.  This will propagate as a systematic uncertainty through to 
the other energies (see equation~\ref{dop2}).

Uncertainties in $v_{\mathrm{O}^+}$ have the same effect as 
uncertainties in the wavelength calibration, which we deduce is accurate to 
0.5 km/s, consistent with the scatter of individual blue H$^+$ line 
velocities about the mean.  Thus, for the remainder of the Orion analysis, 
we adopt $\delta v_{\mathrm{O}^+} = 0.5$~km~s$^{-1}$, noting that this 
subsumes errors in the wavelength calibration of the blue lines.
For the red lines there is an additional systematic error of order 0.15 
km~s$^{-1}$ relating to the alignment of the separate red and blue echelle 
spectra into the same velocity/wavelength system.
The next issue is the accuracy of the wavelength calibration for individual 
lines.
To assess this, we note that there are numerous lines that 
are duplicated in neighbouring echelle orders of the same spectrum.  By
comparing the duplicate measurements of these lines' wavelengths, 
we find a residual characteristic order-to-order difference of 
0.65 km s$^{-1}$, independent of measurement uncertainty.
In practice, we feel that measurements of the red [O~{\sc ii}] lines, which all 
appear in the centre of an order, have a systematic uncertainty of less than 
half this, $\sim0.3$~km~s$^{-1}$.
Taken all together, the combined systematic error is of order 
0.7~km~s$^{-1}$, which corresponds to 
$\delta E_{^2\mathrm{D}_{3/2}-^2\mathrm{P}_{3/2}} = 0.03$.
Therefore, the separation $E_{^2\mathrm{D}_{3/2}-^2\mathrm{P}_{3/2}}$ is 
properly quoted as $13637.34\pm0.01\pm0.03$.  This is significantly
different than what has been previously adopted (see
Table~\ref{tbl1}).

On the other hand, for the small splittings, the systematic effects are tiny 
compared to the model uncertainties, and so are listed below as 0.00.
As anticipated, we find that the value of the splitting
$E_{^2\mathrm{P}_{3/2}-^2\mathrm{P}_{1/2}}$ of the $^2$P term,
$2.02\pm0.01\pm0.00$, is in close agreement with that found
by \cite{der85}, $2.00\pm0.03$ (largely adopted by
\cite{eri87} and \cite{mar93}).

The splitting $E_{^2\mathrm{D}_{5/2}-^2\mathrm{D}_{3/2}}$ of the $^2$D
term is $19.80\pm0.01\pm0.00$, closely consistent with the
value $19.79\pm0.03\pm0.03$ obtained in \S~\ref{bow} from only the
blue lines of planetary nebulae.  We conclude that the splitting of
the $^2$D term should be revised from the value $20.1\pm0.1$
found by \cite{der85} and (largely) adopted by \cite{eri87} and
\cite{mar93}.

These energy differences allow us to calculate the energies of all four 
energy levels (see Table~\ref{tbl1}).
The model was rewritten with these
energies as the parameters in order to track the effects of covariant
changes in this non-linear model of the air wavelengths, and thus
provide the appropriate 
one-dimensional marginal confidence intervals reported
in Table~\ref{tbl1}.
The errors reported are again the model fitting uncertainty and the 
systematic error.  The latter reflects the propagation of the error in
$E_{^4\mathrm{S}_{3/2}-^2\mathrm{D}_{5/2}}$ plus the systematic error in 
 $E_{^2\mathrm{D}_{3/2}-^2\mathrm{P}_{3/2}}$ for the highest two energies

These energy levels allow for the calculation of the air
wavelengths of lines in the UV in addition to those in
Table~\ref{tblvel}.
From $E_{^4\mathrm{S}_{3/2}-^2\mathrm{P}_{3/2}}$
$E_{^4\mathrm{S}_{3/2}-^2\mathrm{P}_{1/2}}$, we calculate 
$\lambda2470.347\pm0.001\pm0.003$ and $\lambda2470.223\pm0.001\pm0.003$, 
respectively.

Because of the presence of systematic errors, the above analysis was carried 
out with equal weights for all lines.  However, we have repeated the 
analysis using a weighted fit based on the uncertainties found in measuring 
each line. 
$E_{^2\mathrm{D}_{5/2}-^2\mathrm{D}_{3/2}}$ increased by only 0.001, 
 $E_{^2\mathrm{D}_{3/2}-^2\mathrm{P}_{3/2}}$ decreased by 0.015 (well within 
the systematic uncertainty) and $E_{^2\mathrm{P}_{3/2}-^2\mathrm{P}_{1/2}}$
increased by 0.007 (recorded in Table~\ref{tbl1} as 0.01 systematic 
uncertainty).

\section{Comparison to Recent Work of Sharpee et al.}

\citet{sha04} have observed the four red 
lines ($\lambda\lambda$7319, 7320, 7330, 7331 \AA) in sky 
spectra using the high-resolution echelle spectrograph (HIRES) on Keck I.
In their determination of the O~{\sc ii} $2p^3$ energy levels, they also
make use of data-sets from nebulae \citep{bow60,der85,bal00,sha03}.  The 
nebular data are given lesser weights (0.018, 0.16, 1.00, 1.00, respectively) 
than the HIRES data-set (10.00).
 Velocity corrections for data-sets of three nebulae 
\citep{der85,bal00,sha03} were found using a weighted least-squares 
analysis similar to what was done in \S~\ref{bow};
\citet{bow60} is an average of multiple nebulae and was assumed to have
$\left<v_{\mathrm{O}^+}\right> = 0$.
From these five data-sets, they obtain splittings of $19.810\pm0.006$ 
and $2.010\pm0.005$ for the $^2$D and $^2$P terms, respectively -- both 
in excellent agreement with our results.

The $^2$D$_{3/2}-^2$P$_{3/2}$ separation, $13637.403\pm0.004$, is quoted 
here with their
1$\sigma$ model fitting uncertainty.
Even though for the HIRES data one knows $v_{\mathrm{O}^+} = 0$, there is 
still a 
systematic uncertainty because of the wavelength calibration which was 
accomplished through the OH Meinel band.
From their stated ``statistical scatter'' of 0.5 km~s$^{-1}$ we judge that 
the systematic error in $E_{^2\mathrm{D}_{3/2}-^2\mathrm{P}_{3/2}}$ is
no larger than
0.023 cm$^{-1}$ (see equation~\ref{dop2}), and 0.012~\AA\ in the wavelengths 
of the red lines.
Based on the model-fitting uncertainties alone, our value 
$13637.34\pm0.01\pm0.03$ might seem significantly 
different, but the values are in fact consistent when one accounts for the 
systematic uncertainties.
If we adopted the \citet{sha04} value of 
$E_{^2\mathrm{D}_{3/2}-^2\mathrm{P}_{3/2}}$, the velocities in 
Table~\ref{tblvel} 
would be more consistent for 37W, but correspondingly less consistent for 
the other two positions.

To complete their set of  O~{\sc ii} $2p^3$ energy levels, the blue 
lines from three nebulae \citep{bow60,bal00,sha04} were used to deduce 
$E_{^4\mathrm{S}_{3/2}-^2\mathrm{D}_{5/2}} = 26810.76\pm0.08$
(no systematic error was estimated).  
This energy difference compares well with our result 
using 23 planetary nebulae, $26810.77\pm0.03\pm0.03$.

The UV wavelengths depend on both $E_{^4\mathrm{S}_{3/2}-^2\mathrm{D}_{5/2}}$ 
and $E_{^2\mathrm{D}_{3/2}-^2\mathrm{P}_{3/2}}$, and so the
\citet{sha04} values 
$\lambda2470.343\pm0.005$ and $\lambda2470.220\pm0.005$ are consistent with 
ours though slightly less accurate.

\section{Conclusions}\label{conc}

From a detailed analysis of Orion nebula and planetary 
nebulae data, we find new energies for the ground configuration.
Both model-fitting uncertainties and systematic uncertainties are 
presented.
This work confirms the utility of astrophysical measurements in 
determining accurate energies for the O~{\sc ii} $2p^3$ ground configuration.
As a by-product, we determine a revised set of air wavelengths of the 
[O~{\sc ii}] visible lines (see Table~\ref{tblvel}) and the UV lines (see 
\S~\ref{con}).
As \citet{sha04} point out, there are revisions required in the standard NIST 
values.

Using these revised wavelengths, it is now possible to constrain 
the velocities
of all three oxygen zones (O$^0$,O$^+$,O$^{++}$), and in turn, the spatial and 
physical origin of oxygen permitted lines can be inferred \citep{bla03b}.

\acknowledgements 
This work was supported by the Natural Sciences and Engineering
Research Council of Canada.  Line wavelengths were obtained from the
 Atomic Line List\footnote{Atomic Line List v2.04 is available at:
http://www.pa.uky.edu/\~{}peter/atomic/.} maintained by P.~A.~M.~van~Hoof.
The authors wish to thank J.~A.~Baldwin for obtaining the spectra, and
G.~J.~Ferland and E.~M.~Verner for their comments on this paper.


\clearpage

\clearpage

\clearpage

\clearpage
\begin{deluxetable}{rrrrrrrr}
\rotate
\tabletypesize{\scriptsize}
\tablecolumns{8}
\tablewidth{0pc}
\tablecaption{O~{\sc ii} ground configuration energy levels (cm$^{-1}$)\label{tbl1}}
\tablehead{
\colhead{Designation} & \colhead{Bowen} & \colhead{De Robertis} & 
\colhead{Level} & \colhead{Eriksson} & \colhead{Martin et al.} &
\colhead{This Work\tablenotemark{a}} & \colhead{Level}\\
\colhead{} & \colhead{(1955)} & \colhead{et al. (1985)} & \colhead{Difference} & \colhead{(1987)} &
\colhead{(1993)} & & \colhead{Difference\tablenotemark{a}}}
\startdata
$^4$S$_{3/2}$ & 0.0 & 0.0 & & 0.00 & 0.00 & 0.00 & \\
$^2$D$_{5/2}$ & 26810.7 & 26810.5 & {\raisebox{1.5ex}[0pt]{26810.5}} & 26810.52 & 26810.55 & 
%
$26810.77\pm0.03\pm0.03$ & {\raisebox{1.5ex}[0pt]{$26810.77\pm0.03\pm0.03$}}\\
$^2$D$_{3/2}$ & 26830.5 & 26830.6 & {\raisebox{1.5ex}[0pt]{$20.1\pm0.1$}} 
& 26830.57 & 26830.57 & $26830.57\pm0.01\pm0.03$ & 
{\raisebox{1.5ex}[0pt]{$19.80\pm0.01\pm0.00$}}\\
$^2$P$_{3/2}$ &  & 40468.1 & {\raisebox{1.5ex}[0pt]{13637.5}} & 40467.69 
%
& 40468.01 & $40467.91\pm0.02\pm0.05$ & {\raisebox{1.5ex}[0pt]{$13637.34\pm0.01\pm0.03$}}\\
$^2$P$_{1/2}$ & {\raisebox{1.5ex}[0pt]{40468.3}} & 40470.1 & 
{\raisebox{1.5ex}[0pt]{$2.00\pm0.03$}} & 40469.69 & 40470.00 & 
$40469.93\pm0.02\pm0.05$ & {\raisebox{1.5ex}[0pt]{$2.02\pm0.01\pm0.01$}}\\
\enddata
\tablenotetext{a}{uncertainties are model-fitting and systematic, 
respectively} \end{deluxetable}

\clearpage
\begin{deluxetable}{lrrrr}
\tabletypesize{\scriptsize}
\tablecolumns{45}
\tablewidth{0pc}
\tablecaption{Determination of $^2$D$_{5/2}$ energy from planetary nebulae data \label{tablepn}}
\tablehead{
\colhead{Nebula} & 
\colhead{$^4$S$_{3/2}$-$^2$D$_{3/2}$ (\AA)} & 
\colhead{$^4$S$_{3/2}$-$^2$D$_{5/2}$ (\AA)} & \colhead{Weight} & 
\colhead{Reference} }
\startdata
NGC 2440 & 3726.00 & 3728.69 & 0.25 & \citet{hyu98}\\
NGC 6543 &
3726.06 & 3728.81 & 1.0 & \citet{hyu00}\\
NGC 6567 &
3726.25 & 3728.97 & 0.5 & \citet{hyu93}\\
NGC 6572 &
3726.01 & 3728.76 & 1.0 & \citet{hyu94b}\\
NGC 6741 &
3726.15 & 3728.90 & 1.0 & \citet{hyu97b}\\
NGC 6790 &
3726.00 & 3728.74 & 1.0 & \citet{all96}\\
NGC 6818 &
3726.32 & 3728.79 & 0.0 & \citet{hyu99a}\\
NGC 6884 &
3726.02 & 3728.80 & 0.5 & \citet{hyu97a}\\
NGC 6886 &
3726.05 & 3728.78 & 1.0 & \citet{hyu95a}\\
NGC 7009major &
3725.97 & 3728.71 & 1.0 & \citet{hyu95b}\\
NGC 7009minor &
3726.14 & 3728.95 & 0.25 & \citet{hyu95c}\\
NGC 7662 &
3725.94 & 3728.69 & 1.0 & \citet{hyu97c}\\
IC 351 &
3726.04 & 3728.81 & 1.0 & \citet{fei96}\\
IC 418 &
3726.04 & 3728.80 & 1.0 & \citet{hyu94a}\\
IC 2149 &
3725.99 & 3728.76 & 1.0 & \citet{fei94}\\
IC 2165 &
3726.18 & 3728.96 & 0.5 & \citet{hyu94d}\\
IC 4634 &
3726.27 & 3729.02 & 1.0 & \citet{hyu99b} \\
IC 4846 &
3726.17 & 3728.93 & 1.0 & \citet{hyu01b}\\
IC 4997 &
3725.97 & 3728.71 & 1.0 & \citet{hyu94c}\\
IC 5117 &
3726.03 & 3728.75 & 0.5 & \citet{hyu01c}\\
IC 5217 &
3726.04 & 3728.77 & 1.0 & \citet{hyu01a}\\
BD +30 3639 &
3726.01 & 3728.77 & 1.0 & \citet{all95}\\
Hubble 12 &
3726.04 & 3728.80 & 1.0 & \citet{hyu96}\\
Hu 1-2 &
3726.07 & 3728.83 & 1.0 & \citet{pot03}\\
\enddata
\end{deluxetable}

\clearpage
\begin{deluxetable}{ccccc}
\tabletypesize{\scriptsize}
\tablecolumns{5}
\tablewidth{0pc}
\tablecaption{Best-fit line parameters for visible wavelength 
transitions.
\label{tbl2}} 
\tablehead{ \colhead{Transition}  & \colhead{Position} &
\colhead{Measured Wavelength (\AA)} & \colhead{FWHM (km s$^{-1}$)} &
\colhead{Reddening corrected line strength} \\
\colhead{} & \colhead{} & \colhead{} & \colhead{} & \colhead{($10^{-13}$ erg 
s$^{-1}$ cm$^{-2}$)} 
}
\startdata
$^4$S$_{3/2}-^2$D$_{3/2}$  & 1SW & $3726.072\pm0.001$ & $16.8\pm0.2$ & 
$426\pm6$ \\
 & x2 & $3726.078\pm0.001$ & $   18.0\pm0.3$ & $273\pm4$ \\
 & 37W & $3726.060\pm0.002$ & $   15.3\pm0.3$ & $466\pm9$ \\
\cline{2-5}
$^4$S$_{3/2}-^2$D$_{5/2}$  & 1SW & $3728.824\pm0.001$ & $17.3\pm0.2$ & 
$188\pm2$ \\
 & x2 & $3728.829\pm0.002$ & $   19.1\pm0.2$ & $138\pm2$ \\
 & 37W & $3728.811\pm0.002$ & $   15.7\pm0.3$ & $213\pm4$ \\
\cline{2-5}
$^2$D$_{5/2}-^2$P$_{1/2}$  & 1SW & $7319.173\pm0.006$ & $18.0\pm0.6$ & 
$14.7\pm0.5$ \\
 & x2 & $7319.178\pm0.008$ & $   21.3\pm0.8$ & $6.9\pm0.2$ \\
 & 37W & $7319.099\pm0.011$ & $   16.9\pm1.1$ & $10.8\pm0.6$ \\ 
\cline{2-5}
$^2$D$_{5/2}-^2$P$_{3/2}$  & 1SW & $7320.253\pm0.002$ & $17.5\pm0.2$ & 
$42.6\pm0.5$ \\
 & x2 & $7320.250\pm0.003$ & $   19.3\pm0.3$ & $17.1\pm0.2$ \\
 & 37W & $7320.181\pm0.004$ & $   16.9\pm0.4$ & $32.0\pm0.6$ \\
\cline{2-5}
$^2$D$_{3/2}-^2$P$_{1/2}$  & 1SW & $7329.787\pm0.004$ & $17.9\pm0.4$ & 
$21.1\pm0.4$ \\
 & x2 & $7329.798\pm0.006$ & $   23.4\pm0.6$ & $8.6\pm0.2$ \\
 & 37W & $7329.725\pm0.005$ & $   16.0\pm0.5$ & $16.6\pm0.4$ \\
\cline{2-5}
$^2$D$_{3/2}-^2$P$_{3/2}$  & 1SW & $7330.886\pm0.003$ & $18.1\pm0.3$ & 
$24.0\pm0.4$ \\
 & x2 & $7330.885\pm0.004$ & $   19.8\pm0.4$ & $9.5\pm0.2$ \\
 & 37W & $7330.818\pm0.005$ & $   17.7\pm0.5$ & $18.2\pm0.5$ \\

\enddata
\end{deluxetable}

\clearpage
\begin{deluxetable}{crrrrrrrrrr}
\tabletypesize{\scriptsize}
\rotate
\tablecolumns{11}
\tablewidth{0pc}
\tablecaption{Comparison between observed and predicted line ratios
as determined from transition probabilities, $A_{ij}$.\label{tblrto} } 
\tablehead{ \colhead{Transition}  & \multicolumn{2}{c}{1SW} &
\multicolumn{2}{c}{x2} & \multicolumn{2}{c}{37W} &
\multicolumn{2}{c}{\citet{zei87}} & \multicolumn{2}{c}{\citet{wie96}} \\
\colhead{} & \colhead{Flux} & \colhead{Ratio} & \colhead{Flux} & 
\colhead{Ratio} & \colhead{Flux} & \colhead{Ratio} & \colhead{$A_{ij}$} &
\colhead{Ratio} & \colhead{$A_{ij}$} & \colhead{Ratio} }
\startdata
$^2$D$_{5/2}-^2$P$_{1/2}$  ($\lambda$7319) & $14.7\pm0.5$ & &  $6.9\pm0.2$ & & $10.8\pm0.6$ & & 0.0563 & & 
$0.0519\pm0.0052$ & \\ $^2$D$_{3/2}-^2$P$_{1/2}$  ($\lambda$7330) &
$21.1\pm0.4$ & {\raisebox{1.5ex}[0pt]{$0.70\pm0.03$}} &
$8.6\pm0.2$ & {\raisebox{1.5ex}[0pt]{$0.80\pm0.03$}}  &
$16.6\pm0.4$ & {\raisebox{1.5ex}[0pt]{$0.65\pm0.04$}}  &
0.0941 & {\raisebox{1.5ex}[0pt]{0.598}} &
$0.0867\pm0.0087$ & {\raisebox{1.5ex}[0pt]{$0.599\pm0.085$}}  \\

$^2$D$_{5/2}-^2$P$_{3/2}$  ($\lambda$7320) & $42.6\pm0.5$ & & $17.1\pm0.2$ & & $32.0\pm0.6$ & & 0.1067  & & 
$0.0991\pm0.0099$ & \\ $^2$D$_{3/2}-^2$P$_{3/2}$  ($\lambda$7331) &
$24.0\pm0.4$ & {\raisebox{1.5ex}[0pt]{$1.77\pm0.04$}} &
$9.5\pm0.2$ & {\raisebox{1.5ex}[0pt]{$1.80\pm0.04$}} &
$18.2\pm0.5$ & {\raisebox{1.5ex}[0pt]{$1.76\pm0.06$}} &
0.0580 & {\raisebox{1.5ex}[0pt]{1.84}} &
$0.0534\pm0.0053$ & {\raisebox{1.5ex}[0pt]{$1.86\pm0.26$}} \\

\enddata
\end{deluxetable}

\clearpage
\begin{deluxetable}{ccccccrrr}
\tabletypesize{\scriptsize}
\tablecolumns{9}
\tablewidth{0pc}
\tablecaption{Line velocities in Orion using recent wavelength tabulations. 
\label{tblvel}}

 
\tablehead{
\colhead{} & \multicolumn{3}{c}{Air $\lambda$ (\AA)}
& \colhead{} & \colhead{} & 
\multicolumn{3}{c}{Velocity\tablenotemark{d}\  (km s$^{-1}$)} \\
\cline{2-4} \cline{7-9} \\
\colhead{Transition}  & \colhead{Martin\tablenotemark{a}} & 
\colhead{Eriksson\tablenotemark{b}} & \colhead{This work\tablenotemark{c}} &
\colhead{Position} &
\colhead{Observed $\lambda$ (\AA)} & 
\colhead{Martin} & \colhead{Eriksson} & \colhead{This work} \\
\colhead{(1)} & \colhead{(2)} & \colhead{(3)} & \colhead{(4)} & 
\colhead{(5)} & \colhead{(6)} &\colhead{(7)} & \colhead{(8)} & \colhead{(9)}

}

\startdata

{$^4$S$_{3/2}-^2$D$_{3/2}$} & {3726.032} &  {3726.032} & {3726.032} &
1SW & 3726.072 & $3.2\pm0.1$ & $3.2\pm0.1$ & $3.2\pm0.1$ \\
 & & & $\pm0.001$ & x2 & 3726.078 & $   3.7\pm0.1$ & $3.7\pm0.1$ & 
$3.7\pm0.1$ \\
 & & & $\pm0.004$ & 37W & 3726.060 & $   2.3\pm0.2$ & $2.3\pm0.2$ & 
$2.3\pm0.2$ \\ \cline{5-9} 
{$^4$S$_{3/2}-^2$D$_{5/2}$} & {3728.815} &  {3728.819} & {3728.784} &
1SW & 3728.824 & $0.7\pm0.1$ & $0.4\pm0.1$ & $3.2\pm0.1$ \\
 & & & $\pm0.004$ & x2 & 3728.829 & $   1.1\pm0.2$ & $0.8\pm0.2$ & 
$3.6\pm0.2$\\
 & & & $\pm0.004$ & 37W & 3728.811 & $   -0.3\pm0.2$ & $-0.6\pm0.2$ & 
$2.2\pm0.2$\\ \cline{5-9} 
{$^2$D$_{5/2}-^2$P$_{1/2}$} & {7318.92} &  {7319.07} & {7319.073} &
1SW & 7319.173 & $10.4\pm0.2$ & $4.2\pm0.2$ & $4.1\pm0.2$ \\
 & & & $\pm0.009$ & x2 & 7319.178 & $   10.6\pm0.3$ & $4.4\pm0.3$ & 
$4.3\pm0.3$\\ 
 & & & $\pm0.012$ & 37W & 7319.099 & $   7.3\pm0.5$ & $1.2\pm0.5$ & 
$1.1\pm0.5$\\ \cline{5-9} 
{$^2$D$_{5/2}-^2$P$_{3/2}$} & {7319.99} &  {7320.14} & {7320.157} &
1SW & 7320.253 & $10.8\pm0.1$ & $4.6\pm0.1$ & $3.9\pm0.1$ \\
 & & & $\pm0.009$ & x2 & 7320.250 & $   10.6\pm0.1$ & $4.5\pm0.1$ & 
$3.8\pm0.1$ \\
 & & & $\pm0.012$ & 37W & 7320.181 & $   7.8\pm0.2$ & $1.7\pm0.2$ & 
$1.0\pm0.2$\\ \cline{5-9} 
{$^2$D$_{3/2}-^2$P$_{1/2}$} & {7329.67} &  {7329.83} & {7329.699} & 
1SW & 7329.787 & $4.8\pm0.1$ & $-1.8\pm0.1$ & $3.6\pm0.1$ \\
 & & & $\pm0.005$ & x2 & 7329.798 & $   5.2\pm0.2$ & $-1.3\pm0.2$ & 
$4.0\pm0.2$ \\
 & & & $\pm0.012$ & 37W & 7329.725 & $   2.2\pm0.2$ & $-4.3\pm0.2$ & 
$1.1\pm0.2$\\ \cline{5-9}

{$^2$D$_{3/2}-^2$P$_{3/2}$} & {7330.73} &  {7330.91} & {7330.786} &
1SW & 7330.886 & $6.4\pm0.1$ & $-1.0\pm0.1$ & $4.1\pm0.1$ \\
 & & & $\pm0.005$ & x2 & 7330.885 & $   6.3\pm0.2$ & $-1.0\pm0.2$ & 
$4.0\pm0.2$ \\
 & & & $\pm0.012$ & 37W & 7330.818 & $  3.6\pm0.2$ & $-3.8\pm0.2$ & 
$1.3\pm0.2$\\

\enddata
\tablenotetext{a}{\citet{mar93}}
\tablenotetext{b}{\citet{eri87}}
\tablenotetext{c}{uncertainties are model-fitting and systematic, respectively, 
from energy level differences (see Table~\ref{tbl1})}
\tablenotetext{d}{uncertainties are simply from the least-squares Gaussian fit 
for each line.}

\end{deluxetable}

\end{document}